# Study of Cloud Computing in HealthCare Industry

G.Nikhita Reddy, G.J.Ugander Reddy

**Abstract**— In Today's real world technology has become a domiant crucial component in every industry including healthcare industry. The benefits of storing electronically the records of patients have increased the productivity of patient care and easy accessibility and usage. The recent technological innovations in the health care is the invention of cloud based Technology. But many fears and security measures regarding patient records storing remotely is a concern for many in health care industry. One needs to understand the benefits and fears of implementation of cloud computing its advantages and disadvantages of this new technology.

**Index Terms**— Cloud, Cloud Computing, HealthCare, Electronic records, Security, HIPAA, Mobile health.

—————— ◆ ——————

## 1 INTRODUCTION

Cloud computing is one of the most recent revolutionary technologies in world.The applications of Cloud computing is rapidly increasing in day to day life.Today the application of cloud computing is so widespread that it is being used even in the health care industry.As the evolution of cloud computing in health care is occurring at a rapid rate in recent times, we can expect a major part of the healthcare services to move onto the cloud and thereby more focus is laid on providing a cost effective and efficient healthcare service to the people all around the globe.

Despite of a common belief that certain boundaries and security issues of the cloud would hinder the shift, the healthcare industry is taking an initiative to move to these cloud based platforms.Today many doctors and hospitals are moving towards these clouds in order to provide better healthcare services to their patients.

According to the research firm Markets and Markets, The cloud computing market in the health care sector is expected to grow by 2017 to $5.4 billion.Hence from this survey it can be interpreted that the applications of cloud in healthcare is going to be a huge industry in the near future.

## 2 CLOUD COMPUTING

Cloud computing may be defined as the use of computing resources both hardware and software that are delivered as a service over a network most likely the Internet.Cloud consists of three basic service models (IaaS, PaaS, and SaaS).

Infrastructure as a Service (IaaS) provides users with processing, storage, networks, and other computing infrastructure resources

Platform as a Service (PaaS) enables users to deploy applications developed using specified programming languages or frameworks and tools onto the Cloud infrastructure

Software as a Service (SaaS**)** enables users to access applications running on a Cloud infrastructure from various end-user devices (generally through a web browser).

- G.Nikhita Reddy is currently pursuing B.E, CSE second year at Chaitanya Bharathi institute of technology, Osmania Univeristy, Hyderabad. India E-mail: nikhita.reddy@yahoo.com
- G.J.Ugander Reddy is B.E, M.B.A and Founder director at Peridot Technologies, Hyderabad. India E-mail: ugander.reddy@peridot-tech.com

Deployment of a cloud can be done in the following ways:
Private clouds: They are operated solely for one organization only or for an individual person.
Public clouds: They are open to the general public or large industrial groups and are owned and are usually managed by a Cloud service provider.
Hybrid clouds: They combine two or more clouds (private or public) that remain unique entities but are bound together by technology that enables data and application portability.
Community clouds: They feature infrastructure that is shared by several organisations and supports a specific community.

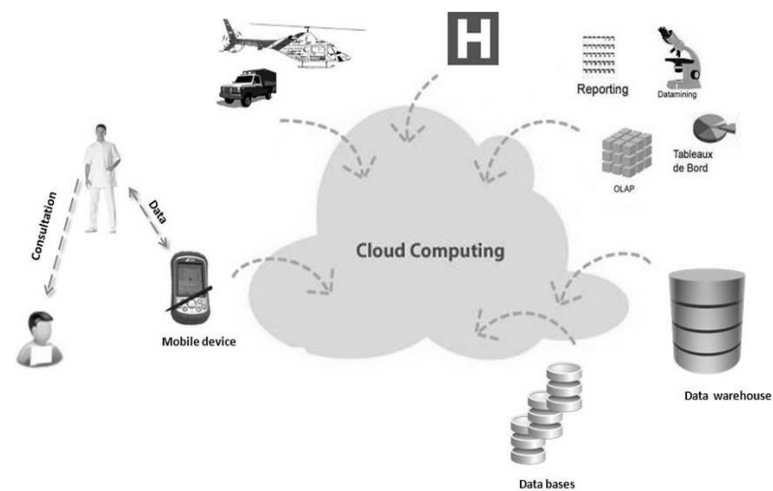

## 3 CLOUD COMPUTING IN HEALTHCARE

In recent years cloud computing technologies are on rise in the health care industries.The demand for cloud technologies in healthcare sector is increasing day by day.The environment in the healthcare sector is changing rapidly than ever before due to the increase in the demand for delivering the most effective medical services for a low cost of money which has increased the compitition between the various healthcare providers.

Hospitals, doctors, research clinics, private and public health care institutions are looking for alternatives in order to increase the efficiency of the services for less money.

If these cloud computing technologies are implemented appropriately, they meet the requirements of all the problems faced by the healthcare industry.Thus these recent cloud technologies provide oppurtunities to healthcare sectors in order to improve their services for the patients, to improve the operational facilities, to share information in an easy way, and to cut down the costs.Hence with the help of cloud computing in healthcare a doctor can access his patients records even if they are miles away.

Thus the use of cloud technologies in healthcare can be proved as a boon for the patients all around the world.

## 4   HIPAA AND ITS REQUIREMENTS

Health Insurance Portability and Accountability Act (HIPAA) was basically designed to protect the privacy of patient's medical health records.

HIPAA does the following:
- It provides the ability to transfer and continue the health insurance coverage for millions of American workers and their families when they change or lose their jobs.
- It reduces fraud and abuse of healthcare.
- It mandates industry-wide standards for health care information on electronic billing and other processes.
- Requires the protection and confidential handling of protected health information.

HIPAA omnibus and the American Recovery and Reinvestment Act (ARRA) requirements demanded everyone in the healthcare industry to begin the movement of patient's records and other data to cloud computing technologies.

Essentially it is estimated that by 2015, all medical professionals with access to appropriate patient records must utilize the electronic medical/health records (EHR'S or EMR'S) or will face penalities.

## 5   EMR'S / EHR'S

An electronic health record (EHR) can be defined as a proper and systematic collection of health information about an individual person or population in an electronic form.It is uaually recorded in a digital format and hence can be easily accesed by a clinical practitioner or can be easily shared between different healthcare centers.Hence by adopting EHR'S storing the health information of the concerned individual becomes easy.

These days adoption of EHR'S is being increasing considerably by the physicians.The biggest advantage of shifting the EHR'S to clouds is that the information about the patients health can be shared by various healthcare facilities, clinical practitioners and others in order to facilitate quick and quality patient treatments.

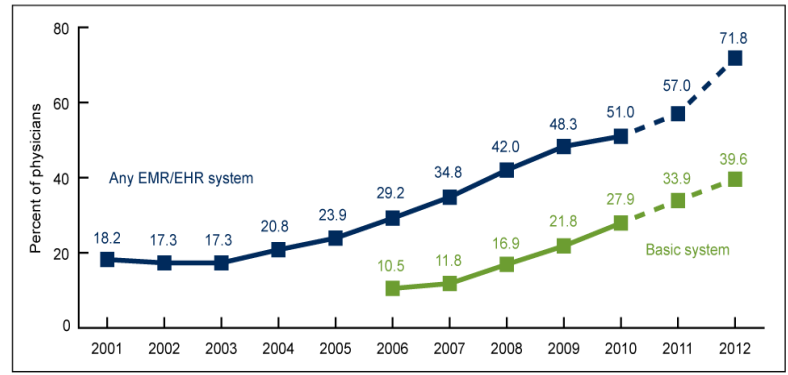

Percentage of office-based physicians with EMR/EHR systems: United States, 2001–2010 and preliminary 2011–2012

The above figure clearly shows the increase in EMR adoptions every year.Even though EHR's can become a huge benefit to the patients, utilization of clouds should meet the requirements some of which may include data reversibility, traceability of access of the patient, confidentiality of the reports, and some security issues.Even the public cloud systems must satisfy the regulations of HIPAA and other electronic record guidelines for their successful utilization.Once if these requirements are met EHR's will become the most authentic way of storing the health information of a person safely and securely.

## 6   BENEFITS OF CLOUD COMPUTING IN HEALTHCARE

There are immense benefits and advantages upon implementation of cloud computing in healthcare industry some of which may include:

I) Mobility of records:

In some cases a person's health information can be required by two or more health institutions in that case by implementation of cloud technologies a person's health information can be easily synchronized and shared at the same time.Hence this improves physicians ability to provide a better health care to the patients.Thus by implementation of cloud technologies a patients information is readily available.

II) Speed:

By using cloud based technologies and services always enable faster and accurate access to all the important information for the healthcare services providers and the history of their patients.

III) Security and Privacy:

By using cloud computing is mainly used for storage of medical records online.With the recent HIPAA update, cloud healthcare service providers are now accountable for HIPAA compliance as a healthcare entities they serve.Thus this includes encryption of data and secure backup of this data which contains the health information of a person, then verifying if the data can be easily regained, and finally security can be improved by using permission based and secured data bases.

IV) Reduction of costs:
By adopting these cloud techniques in healthcare patients, physicians, other medical organizations experience cost savings to a great extent.Since there is no need for these healthcare institutions and doctors to invest huge amounts in hardware infrastructure and their maintainence as these problems are already handled and taken care by the cloud computing providers.According to a recent report by Healthcare Financial Management, says depending on the size and scope of the healthcare institution the savings achieved by utilization of EHR's can amout upto $37 million over the next five year period.

## 7 FEARS OF CLOUD COMPUTING IN HEALTHCARE

Even though implementation of cloud computing in healthcare industry can have several benefits yet there are some fears also in the implantation of this latest technology in a complex field like healthcare.Security of the patient data is still the biggest concern in this area.

I) Privacy and security challenges:
Always the data maintained in the cloud may contain either personal, private or confidential information regarding a person's health status and his health records which ought to be properly safe-guarded inorder to prevent the misuse of this information and their disclosure.Global concers related to data jurisdiction, pivacy of data, security and compliance are having a huge impact in adoption of these cloud technologies by healthcare organizations.

II) Type of cloud:
There has always been a never ending debate on determining the best cloud model (public or private) that would meet the needs of healthcare induatry in a better way.The biggest disadvantage of using a public cloud is that it lacks the control and the security policies required for a health organization, on the other hand a private cloud provides good customization, security, privacy and also possess a high level of internal control of the data.But still because of their better security levels private clouds may have an edge over the public clouds which are being proven to be insecure when comes to handling the information.But still much research is still goin on in order to know which cloud suits the needs of healthcare industry the best.

III) Data Portability:
Another biggest challenge that some of the healthcare organizations face in adopting the cloud technologies' is the concern regarding the ability to transition to another cloud vendor or back to the healthcare organization without interrupting operations or introducing conflicting claims to the data. With traditional IT technologies, the healthcare organization has physical control of systems, services and data. The concern is that if a provider were to suspend its services or refuse access to data, a healthcare organization may suddenly be unable to provide the required healthcare services its patients or customers' wchich creates a problem.

IV) Service Reliability:
The reality is that all the cloud technologies and other enterprise infrastructures will definitely have certain interruptions Of some degree at any certain point of time.Thus the applications must met very high performance,availability and good reliability standards.The day to day growing reliance on distributed network based solutions are only increasing the difficulties and complexities of securing and maintaining the data in these dyanamic environments.

The dependence of this healthcare industry on availability and reliability of information can be a matter of life and death.

Disaster recovery is a component of service reliability that focuses on processes and technology for resumption of applications, data, hardware, communications (such as networking), and other IT infrastructure in case of a disaster.

Performance is another factor which is having a impact on slow adoption of cloud computing in healthcare industries.

## 8 MOBILE HEALTH

Mobile health or mHealth usually uses mobile technologies as the basic elements for health research and delivering the healthcare services.In the future years to come these mobile systems will be able to retrieve and update the data that is sored in the electronic health records by the utilization of latest cloud technologies.

Doctors and clinical physicians are still thinking on utilization of EHR's since they have their own disadvantages in that case mobile health can be a huge plus.

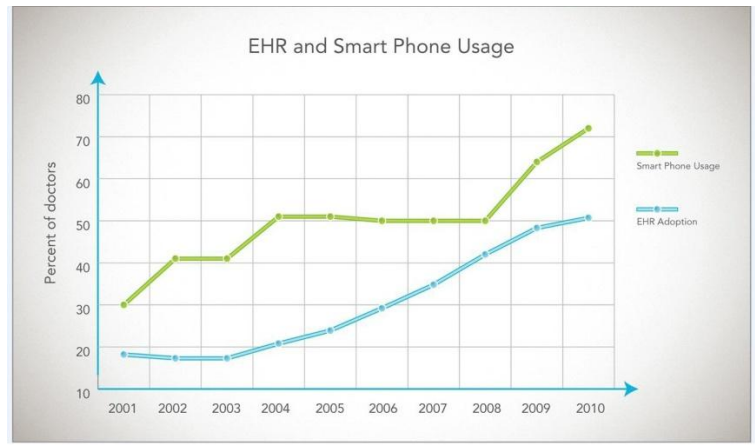

From the above figure it is clear that the percentage of doctors preferring mobile phones tham EMR's/EHR's has always beem more and is also increasing day by day.The mobile health may appeal well to the doctors over EHR's since these EHR's lack the storage of structured data which can be done by using this mobile technologies.Still mobile health records also have some disadvantages regarding the security issues, a lot more research is still going on in this field of mobile health care by utilizing the cloud technologies.

The major advantage of using mobile technologies with cloud technologies is its mobility and easy sharing of the information.

## 9 CONCLUSION

Cloud computing is changing our lives in many ways at a very quick pace.Day by day utilization of cloud computing technologies is increasing in every part of the world.There are several reasons as discussed above for the utilization of cloud technologies in healthcare industry.The cloud computing solutions in healthcare can help the physicians to stay in touch with their patients and examine their health condition effectively at a low cost.There may be some concern regarding the security and other issues of data but still as every problem has a solution in the similar way these issues too will be overcomed one day by man after which utilization of cloud technologies in healthcare industry would result in a new era in the field of healthcare.Every section in the society can access this healthcare by implementation of this technology.

It is always remembered that cloud computing is still a developing technology,which implies that in the future years the services it offers will be greater than our expectations or jush beyond our imagination.